\documentclass[aps,prl,twocolumn,superscriptaddress,showpacs]{revtex4}
\usepackage[dvips]{graphicx}

\begin{document}

\title{ Phonon Coherence and New Set of Sidebands in Phonon-Assisted
Photoluminescence }

\author{Shi-Jie Xiong}
\email{sjxiong@nju.edu.cn}
\address{National Laboratory of Solid State Microstructures and
Department of Physics, Nanjing University, Nanjing 210093, China}
\address{Department of Physics and HKU-CAS Joint Laboratory on
New Materials, The University of Hong Kong, Pokfulam Road, Hong Kong, China}
\author{Shi-Jie Xu}
\email{sjxu@hkucc.hku.hk}
\address{Department of Physics and HKU-CAS Joint Laboratory on
New Materials, The University of Hong Kong, Pokfulam Road, Hong Kong, China}

\begin{abstract}

We investigate excitonic polaron states comprising a local exciton
and phonons in the longitudinal optical (LO) mode by solving the
Schr\"{o}dinger equation. We derive an exact expression for the
ground state (GS), which includes multi-phonon components with
coefficients satisfying the Huang-Rhys factors. The recombination
of GS and excited polaron states gives one set of sidebands in
photoluminescence (PL): the multi-phonon components in the GS
produce the Stokes lines and the zero-phonon components in the
excited states produce the anti-Stokes lines. By introducing the
mixing of the LO mode and environal phonon modes, the exciton will
also couple with the latter, and the resultant polaron states
result in another set of phonon sidebands. This set has a
zero-phonon line higher and wider than that of the first
set due to the tremendous number of the environal modes. The energy
spacing between the zero-phonon lines of the first and second sets
is proved to be the binding energy of the GS state. The common exciton
origin of these two sets can be further verified by a characteristic Fano
lineshape induced by the coherence in the mixing of the LO and the environal modes.

\end{abstract}


\maketitle


It has been recognized that the interactions of excitons with
phonons in light-emitting materials lead to a number of
interesting effects in optical properties, such as the phonon
sidebands in photoluminescence (PL) spectra, the exciton dephasing
and self-trapping \cite{1,2,3,4,5,a5,6,7}. Theoretically, the
sideband structures caused by electron-phonon interactions for
local excitons at deep centers were investigated using the
Huang-Rhys model \cite{8,9,10}. In these theories, an adiabatic
approximation is applied to separate the degrees of freedom of
excitons and phonons. One set of Stokes lines (SLs) in PL,
observed in the experiments, has been very successfully explained by
the theories. However, there are some features in the experimental
spectra associated with the interactions of excitons and phonons,
such as complicated structure containing more than one set of
phonon sidebands, anti-Stokes lines (ASLs), and Fano-like
lineshape, seemingly beyond the adiabatic approximation. Recently,
the fabrication of nanostructures has renewed the study of this
field, as the confinement in quantum dots much enhances the
interaction between carriers and phonons. Such an enhancement has
been experimentally observed \cite{aa1,aa2}. Fomin {\it et al.}
have proposed a nonadiabatic approach to interpret the unusual
phonon sidebands observed in quantum dots \cite{dev}. Verzelen
{\it et al.} have indicated that the exciton-phonon bound states
as a whole, called excitonic polarons, and the decaying of their
phonon components into phonon thermostat have to be put under
consideration in cases of strong exciton-phonon interactions
\cite{11}.

In this Letter, we present a unified theory beyond the adiabatic
approximation to explain complicated structures of phonon
sidebands in PL. The exact eigenstates and eigenenergies of
excitonic polarons are obtained by solving the Schr\"{o}dinger
equation of an exciton interacting with LO mode. The obtained
ground state (GS) contains multi-phonon components with
coefficients strictly satisfying the Huang-Rhys factors, which gives the
SLs including a ZPL. Meanwhile the zero-phonon components of the excited states
can produce the ASLs. These SLs and ASLs form one set of sidebands.
By introducing the mixing of LO mode and environal modes,
another set emerges. The energy shift between two sets is
determined by the strength of exciton-LO phonon coupling, and the
coherence in the mode mixing results in specific Fano-like
lineshape.

First, let us consider the interaction between an
exciton and a LO mode
\begin{equation}
\label{ham}
 H = \epsilon_{0} a^{\dag} a
  + \hbar \omega_{0} b^{\dag}_{0} b_{0}
  + V_0 a^{\dag} a (
b^{\dag}_{0} + b_{0} ) ,
\end{equation}
where $a$ and $b_{0}$ are annihilation operators for exciton and
for LO phonon, $\epsilon_{0}$ and $\hbar \omega_{0}$ are their
energies, respectively, and $V_0$ is the coupling strength between
exciton and phonon. Here we regard the exciton as one
quasiparticle, supposing that it has a large binding energy.


For Hamiltonian (\ref{ham}), the $m$th eigen-wavefunctions can be
written as
\begin{equation}
\label{ent}
  \Psi_m = \sum_{ n} c_{m;  n  }
   \frac{a^{\dag} (b^{\dag}_{0} )^{n}}{\sqrt{n!}} | 0 \rangle ,
\end{equation}
where $n$ is the number of phonons of the LO mode in the
corresponding component, $|0 \rangle $ is the vacuum, and the
coefficients satisfy the following iteration relations
\begin{equation}
\label{eigen}
 [E_m -(\epsilon_0 + n \hbar \omega_0 )] c_{m; n}
= V_0 \sqrt{n}
  c_{m;n-1 }
 + V_0 \sqrt{n+1}
  c_{m; n+1 }.
\end{equation}
For the GS ($m=0$), $E_0$ is below $\epsilon_0$, and coefficients
are related with each other by
 \[
   c_{0;1} = \frac{(E_0 -\epsilon_0) c_{0;0}}{V_0},
   \]
   \[
   c_{0;2} =
   \frac{(E_0 -\epsilon_0-\hbar\omega_0)c_{0;1}-V_0
   c_{0;0}}{\sqrt{2} V_0}
   \]
   \[
   = \frac{(E_0-\epsilon_0)c_{0;1}}
   {\sqrt{2} V_0} +\frac{-\hbar \omega_0 (c_{0;1}/c_{0;0})-V_0}
   {\sqrt{2}V_0} c_{0;0},
   \]
   \[
   \ldots,
   \]
   \[
c_{0;n+1} =
   \frac{(E_0 -\epsilon_0-n\hbar\omega_0)c_{0;n}-V_0
   \sqrt{n}c_{0;n-1}}{\sqrt{n+1} V_0} \]
   \begin{equation}
   = \frac{(E_0-\epsilon_0)c_{0;n}}
   {\sqrt{n+1} V_0} +\frac{-n\hbar \omega_0 (c_{0;n} /c_{0;n-1})
   -V_0 \sqrt{n}}
   {\sqrt{n+1} V_0} c_{0;n-1}.
   \end{equation}
From these relations, we can find the exact expressions of the GS energy
\begin{equation}
 \label{gs}
  E_0 = \epsilon_0 - \Delta_0
  \end{equation}
with $\Delta_0 \equiv \frac{V_0^2}{\hbar \omega_0}$ being the
binding energy of the excitonic polaron, and the coefficients
\begin{equation}
   c_{0;n}= \frac{(-1)^n S^{n/2} }{e^{S/2} \sqrt{n!} },
\end{equation}
with $S$ being the Huang-Rhys factor
$
 S=\frac{V_0^2}{\hbar^2 \omega_0^2 } \equiv \frac{\Delta_0}{\hbar \omega_0}.
$ Considering that the $n$th component in GS $\Psi_{m=0}$ is
associated with $n$ phonons, $|c_{0;n}|^2$ gives the intensity of
a phonon sideband at energy $E_0 - n \hbar \omega_0 $.
This is as predicted by the Huang-Rhys theory.

\begin{figure}[h]
\includegraphics[width=8.6cm]{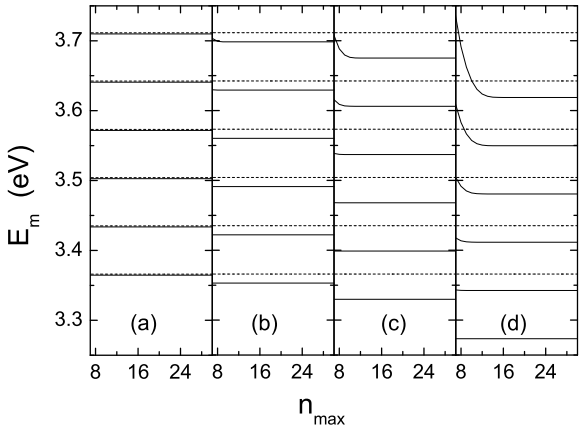}
\caption { Several low-lying eigenenergies $E_m$ as functions of
$n_{\text{max}}$, the maximum number of phonons included in the
calculation. $\epsilon_0=3.366$eV, $\hbar\omega_0 = 0.069068$eV,
(a) $V_0=0.01$eV, (b) $V_0=0.03$eV, (c) $V_0=0.05$eV, (d)
$V_0=0.08$eV. The dotted lines represent the corresponding values
of $\epsilon_0 + m\hbar \omega_0 $ for comparison. } \label{fig1}
\end{figure}

For excited states $\Psi_m$ with $m > 0$, exact analytical
expressions can not be obtained in this way. However, from a
numerical diagonalization shown in Fig. 1 one can see that for
both cases of weak and strong coupling the eigenvalues can be
expressed as
\begin{equation}
 E_m = \epsilon_0 +m\hbar \omega_0 - \Delta_0.
 \end{equation}
From this the coefficients in the eigenfunctions can be obtained
by the following iteration relations
\begin{equation}
  c_{m;n+1} = -\frac{  c_{m;n} }{ \sqrt{n+1} V_0 D_{m;n}}, \text{ for
  } n=0,1,2,\ldots,
  \end{equation}
where
\[
  D_{m;n+1}=\frac{1}{ \Delta_0 -(m-n-1) \hbar \omega_0
  - (n+1) V_0^2D_{m;n}}, \]
  with the initial value
    \[
  D_{m;0} = \frac{1}{ \Delta_0 -m\hbar \omega_0},
  \]
and by the normalization. Although the eigenenergy of state
$\Psi_m$ with $m>0$ is higher than $E_0$, it should have a finite
occupation rate as long as the photon energy of the excitation
laser is high enough \cite{15}. In order to calculate the PL
spectrum of the polaron states, we must know the population of
each state. For simplicity, we assume that all the states with
energies less than the laser energy $E_{\text{laser}}$ have the
same occupation rate. The intensity of PL at photon energy
$\epsilon $ is proportional to
\begin{equation}
\label{rro} \rho (\epsilon)= \sum_{m}  \sum_n |c_{m;n}|^2
\theta(E_{\text{laser}}-E_m) \delta(\epsilon - E_m + n \hbar
\omega_0 ).
\end{equation}

The luminescence peaks produced by the GS $\Psi_{0}$ consist of a
ZPL at energy $E_0 = \epsilon_0-\Delta_0 $ and a number of SLs at
energies $E_0 - n\hbar\omega_0$. This ZPL originates from the
zero-phonon component ($c_{0;0}$) of the GS state while the
remaining lines are from the components ($c_{0;n>0}$) with more
than one phonon. The intensities of the SLs relative to the ZPL
follow a Poisson distribution predicted by the Huang-Rhys theory.
Other eigenstates $\Psi_{m>0}$ will produce complex sideband
structures, which is not dealt with in the Huang-Rhys theory.
Radiative recombination of these states will intensify the ZPL
peak by accompanying with simultaneous emission of $m$ LO phonons.
When they emit $n>m$ phonons in the radiative recombination, the
intensity of the SL at $E_m -n\hbar \omega_0$ will be enhanced.
Meanwhile, the component $c_{m;n<m}$ of state $\Psi_m$ may produce
the ASL peak at $E_m -n\hbar \omega_0$ above the ZPL, together
with the emission of $n(<m)$ phonons. The relative intensity of
this ASL is $ \rho_{m;m-n} /\rho_{m;m} = \frac{S^n (m)!}{n!^2
(m-n)!}$.

It is known that besides LO mode, there are a large number of bath
phonon modes in solids. Although these bath modes do not directly
interact with the exciton, they may have crucial influence on
phonon components of the polaron states. This influence can be
accounted for using the following Hamiltonian of the mode mixing
\begin{equation}
  \label{h1}
  H_1=
 \sum_{\lambda} g
  ( b^{\dag}_0b_{\lambda } +
\text{H.c.}) +\sum_{\lambda}
  \hbar \omega_{\lambda}(b^{\dag}_{\lambda} b_{\lambda } +\frac{1}{2}),
\end{equation}
where $\lambda$ is the mode index of bath phonons, and $g$ is the strength of the mixing. For simplicity, $g$ is assumed to be the same for all bath modes and the dispersion relation $\omega_{\lambda}$ is continuous with a constant density of states between energies 0 and $W$. For a pure harmonic phonon system without exciton, the mixing
terms should vanish as all the phonon modes are orthogonal to each
other. The formation of the excitonic polaron states, however, violates the
orthogonality between the LO and bath modes, and thus creates
the mixing terms. It is this mixing that makes the LO phonons
in the polaron states $\Psi_m$ decays into the continuum of
bath modes. From Hamiltonian (\ref{h1}), the orthogonal modes
become
\begin{equation}
  \label{mmode}
   b_l= \frac{ \hbar \omega_l -\hbar \omega_0}{\sqrt{
   g^2+(\hbar\omega_0 -\hbar \omega_l)^2}} b_{\lambda}
   +\frac{g}{\sqrt{g^2 +(\hbar \omega_0 -\hbar
   \omega_{l})^2}} b_0,
   \end{equation}
with frequency $\omega_l$ determined by solving the equation
\begin{equation}
  \hbar \omega_l= \hbar \omega_0 + \sum_{\lambda }
  \frac{g^2}{ \hbar \omega_l -\hbar \omega_{\lambda}}.
  \end{equation}
The resultant set of orthogonal modes $\{l\}$ will couples with the
exciton at a strength
\begin{equation}
  V_l = \frac{ g}{ \sqrt{g^2 +(\hbar \omega_0
  -\hbar \omega_l)^2}} V_0.
  \end{equation}
Figure 2 shows the phonon frequency dependence of $V_l$. It can be
seen that most of the modes $\{l\}$ are still weakly coupled to
the exciton except for those modes with frequencies $\omega_l \sim
\omega_0$. As mentioned earlier, this results in not only the
broadening of the peaks in the first set but also the creation of
the second set of the phonon sidebands.

\begin{figure}[ht]
\includegraphics[width=9.6cm]{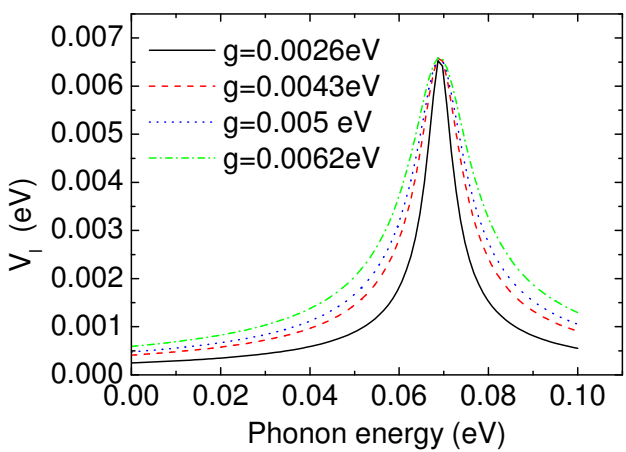}
\caption { Coupling strength $V_l$ as a function of phonon
frequency. } \label{fig2}
\end{figure}

To give a clarification, we can divide the modes $\{l\}$ into two
groups: $P_1$ includes modes with $\omega_l \sim \omega_0$ which
strongly couple with the exciton; $P_2$ includes the remaining
modes weakly coupled to the exciton. Because the modes in $P_1$
with frequencies in the vicinity of $\omega_0$ have coupling
strength of about $V_0$ with the exciton, the formed polaron
states will produce the first set of phonon sidebands including
broadened ZPL, SLs and ASLs, as already discussed above. It should
be noted that the ZPL is centered at the GS eigenenergy $E_0$
which lows with increasing $g$ due to an increase of the mode
number in group $P_1$ as seen from Fig. 2. Now we turn to discuss
the modes in $P_2$. For a mode $l$ in this group, it has a weak
coupling with the exciton, i.e., $V_l\sim 0 $, so the binding
energy of the relevant polaron state is approaching zero. In other
words, the eigenenergy of the polaron state is very close to the
bare exciton energy $\epsilon_0$. Thus, the ZPL produced by the
zero-phonon component in this state almost located at
$\epsilon_0$, higher than the ZPL of the first set by an energy
about $\varepsilon_0-E_0$. The intensity of the ZPL in the second
set should be much stronger than that of the ZPL in the first set
since the ZPL intensity is proportional to the number of involved
phonon modes. Because of the evident dispersion of the bath modes
in group $P_2$, it can be expected that the ZPL of the second set
is wider than that of the first set. The energetic position of the
one-phonon sideband in the second set ranges from $\epsilon_0$
(the ZPL position) to $\epsilon_0-W$. The transition probability
of this sideband is found to be
\begin{equation}
 \label{fano}
p_{\,l0;1}\propto \frac{(\hbar \omega_l -\hbar \omega_0+g)^2}{g^2
+(\hbar \omega_0 -\hbar \omega_l)^2}.
 \end{equation}
From Eq. (\ref{fano}), it can be easily known that $p_{l0;1}$
reaches its maximum value for $\hbar\omega_l=\hbar\omega_0+g$ and
the minimum for $\hbar\omega_l=\hbar\omega_0-g$. When the phonon
frequency is increased from $\hbar\omega_0-g$ to $\hbar\omega_0+g$
the intensity of the sideband rapidly increases. However, as the
phonon frequency is further increased over $\hbar\omega_0+g$ the
sideband intensity slowly decreases. This results in a typical
Fano asymmetric lineshape. As a matter of fact, one can rewrite
Eq. (\ref{fano}) into a standard Fano lineshape function $\alpha_l
= (\tilde{\epsilon} + 1)^2 /(\tilde{\epsilon}^2+1)$ with
$\tilde{\epsilon}=(\hbar \omega _{l} - \hbar \omega_0)/g$. Here
the Fano asymmetry parameter is the unity \cite{16}. Since the
mixing between LO-mode and $\lambda$-mode (which forms the $l$
mode components in the polaron states, illustrated in Eq.
(\ref{mmode})), introduced by $H_1$, is coherent, an interference
between the optical transitions associated with the two phonon
components in the $l$ mode will occur. That is the origin of Fano
interference in the phonon-assisted luminescence.

\begin{figure}[ht]
\includegraphics[width=8.6cm]{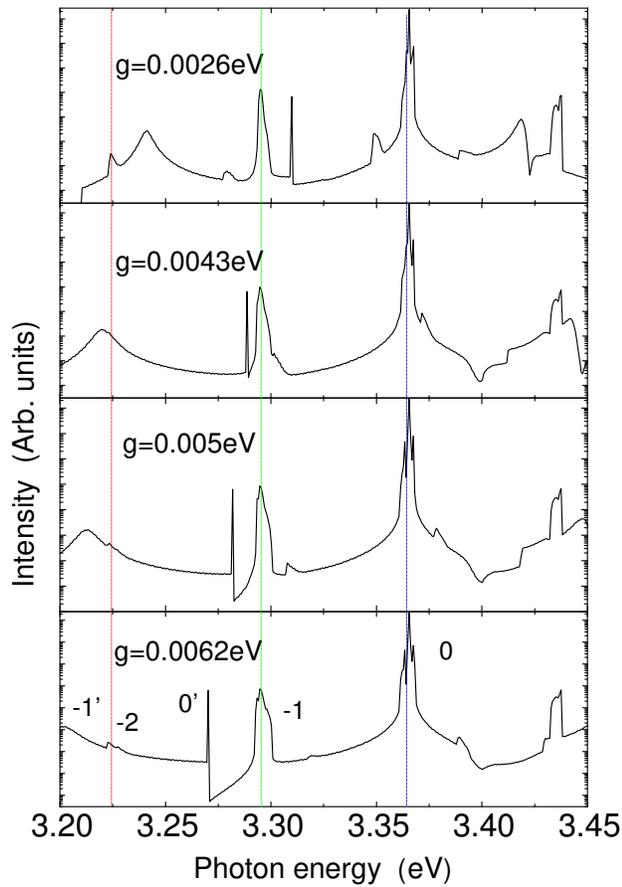}
\caption { Relative intensity as a function of photon energy when
$H_1$ is included. In the simulation 1000 modes uniformly
distributed in a band from 0 to 0.1eV are included. $V_0 =
0.0078$eV. Other parameters are the same as those in Fig. 1. }
\label{fig3}
\end{figure}

By diagonalizing the whole Hamiltonian $H+H_1$, one can calculate
the PL spectrum by
\[
 \rho (\epsilon)= \sum_{m}  \sum_{\{ n_l \}} |c_{m;\{ n_l \}}|^2
\left( \prod_l \alpha_l^{n_l} \right)\theta(E_{\text{laser}}-E_m)
 \]
 \begin{equation}
  \label{ccc}
 \times \delta(\epsilon - E_m + \sum_l
n_l \hbar \omega_l ),
\end{equation}
where $E_m$ and $c_{m;\{ n_l \}}$ are eigenenergy and
eigenfunction of the whole Hamiltonian. In Fig. 3, we plot the PL
spectrum obtained from a numerical calculation. As expected, there
are two ZPL peaks, 0' and 0, belonging to the first and second
set, respectively. In each set, there are several SLs and ASLs.
The SLs are red-shifted from their corresponding ZPL by integer
times $\hbar\omega_0$ while the energy distances of the ASLs from
the ZPL are not integer times $\hbar\omega_0$. It is because that
the energy positions of the ASLs are determined by energies of the
excited polaron states whose energies are not exactly higher than
the GS energy by integer times $\hbar\omega_0$. The Fano
interference features are partially smeared by the summation in
Eq. (\ref{ccc}), but they can still be seen near some peaks. These
results catch the essential nature observed in experiments
[\onlinecite{13}].

In a summary, by reconsidering the exciton-phonon interaction and
decaying of LO phonons we can conclude that there are two sets of
phonon assisted peaks: one is from excitonic polaron states of the
phonon modes with frequencies $\omega_l \sim \omega_0$, and the other is from polaron
states of the remaining bath phonon modes. The latter set was not properly considered
in previous theories and was usually assigned to be the first set because it may include the strongest ``ZPL''
peak. The common exciton origin of these two sets of phonon sidebands can be further
verified by the characteristic energy difference between them
determined by the exciton-LO phonon coupling strength, and by the
specific Fano lineshape reflecting the coherence in the phonon
mode mixing. Together with the experimental observation, the
present theory may give a new and better understanding of the complicated
structures in the phonon-assisted luminescence spectrum.

{\it Acknowledgments} This work was supported in HK by HK RGC CERG
Grant (No. HKU 7036/03P), in Nanjing by National Foundation of
Natural Science in China Grant Nos. 60276005 and 10474033, and by
the China State Key Projects of Basic Research (G20000683).



\begin{references}

\bibitem{1} A. S. Davydov, {\it Theory of Molecular Excitons}, Plenum, New York, 1971.

\bibitem{2} R. S. Knox, in {\it Solid State Physics}, Suppl. 5, edited by F. Seitz and D.
Turnball, Academic, New York, 1963.

\bibitem{3} {\it Polarons and Excitons}. edited by C. G. Kuper and G. D. Whitfield,
Plenum, New York, 1963.

\bibitem{4} E. Peter, J. Hours, P. Senellart, A. Vasanelli, A. Cavanna,
J. Bloch, and J. M. G\'{e}rard, Phys. Rev. B. {\bf 69}, 041307(R) (2004).

\bibitem{5} P. Borri, W. Langbein, U. Woggon, M. Schwab, M. Bayer,
S. Fafard, Z. Wasilewski, and P. Hawrylak,
Phys. Rev. Lett. {\bf 91}, 267401 (2003).

\bibitem{a5} U. Hohenester and G. Stadler, Phys. Rev. Lett. {\bf 92}, 196801 (2004).

\bibitem{6} L. D. Landau, Phys. Z. Sowjetunion {\bf 3}, 664 (1933).

\bibitem{7} T. G. Castner and W. K\"{a}zig, J. Phys. Chem. Solids,
{\bf 3}, 178 (1957).

\bibitem{8} K. Huang and A. Rhys, Proc. Roy. Soc. (London) {\bf A204},
406, (1950).

\bibitem{9} C. B. Duke and G. D. Mahan,
Phys. Rev. {\bf 139}, A1965 (1965).

\bibitem{10} B. Segall and G. D. Mahan, Phys. Rev. {\bf 171}, 935 (1968).

\bibitem{aa1} V. Jungnickel and F. Henneberger, J. Lumin. {\bf
70}, 238 (1996).

\bibitem{aa2} M. Bissiri, G. Baldassarri, H. von H\"{o}gersthal,
A. S. Bhatti, M. Capizzi, A. Frova, P. Frigeri, and S. Franchi,
Phys. Rev. B {\bf 62}, 4642 (2000).

\bibitem{dev} V. M. Fomin, V. N. Gladilin, J. T. Devreese, E. P.
Pokatilov, S. N. Balaban, and S. N. Klimin, Phys. Rev. B {\bf 57},
2415 (1998); V. N. Gladilin, S. N. Klimin, V. M. Fomin, and J. T.
Devreese, {\it ibid.} {\bf 69}, 155325 (2004); V. A. Fonoberov,
E. P. Pokatilov, V. M. Fomin, and J. T. Devreese, Phys. Rev. Lett. {\bf 92}, 127402 (2004).

\bibitem{11} O. Verzelen, R. Ferreira, and G. Bastard, Phys. Rev. Lett. {\bf 88},
146803 (2002).

\bibitem{15} H. Zhao, S. Moehl, S. Wachter, and H. Kalt, Appl. Phys. Lett. {\bf 80},
1391 (2002).


\bibitem{16} U. Fano, Phys. Rev. {\bf 124}, 1866 (1961).

\bibitem{13} S. J. Xu, S. J. Xiong, and S. L. Shi, preprint.

\end{references}
\end{document}